# The Integrated Information Theory needs attention

Azenet Lopez & Carlos Montemayor

# 1      Introduction

The Integrated Information Theory (IIT; Albantakis et al., 2023; Tononi, Albantakis et al., 2022; Ellia et al., 2021) is currently in the spotlight of consciousness science. A survey of scientific papers between 2001 and 2019 includes IIT as one of the four more popular theories (Yaron et al. 2022). Moreover, the massive-scale adversarial-collaborative studies recently conducted by the COGITATE consortium, which constitute a new methodological staple in consciousness research, found substantial confirmation for several IIT predictions (Melloni et al., 2023). Thus, IIT might be one of our current best bets at solving the "easy" problem of consciousness –the problem of finding *which* neural and functional mechanisms give raise to conscious experience (Chalmers, 1996).

But IIT has more to offer. IIT might be our current best bet at a scientific explanation of *phenomenal* consciousness (Block, 1995); thus, it might be our current best bet at solving the infamous hard problem of consciousness –the problem of explaining *why* a given mechanism should give raise to conscious experience. IIT focuses on the distinctively subjective and phenomenal aspects of conscious experience, for which it offers the fundaments of a formal account whose future developments shall explain, in physical and mechanistic terms, why any possible conscious experience *feels* the way it does from the point of view of its subject. At the moment, IIT is well known for its five phenomenological axioms, purportedly capturing the essential properties of any subjectively conscious experience and serving as the starting point of any further theorization. Figure 1 succinctly presents these axioms.

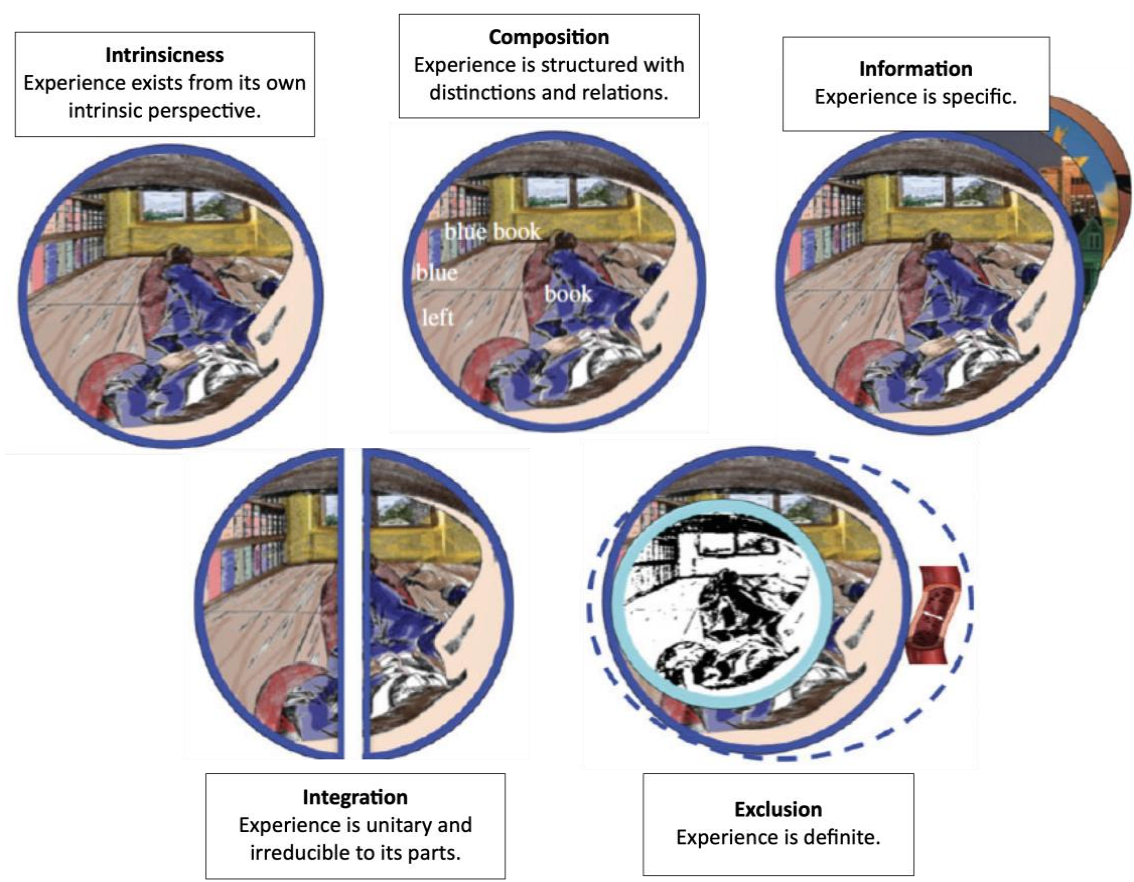

**Fig.1.** IIT's axioms, stating five essential properties of experience.

Adapted from

https://en.m.wikipedia.org/wiki/File:Axioms_and_postulates_of_integrated_information_theory.jpg (CC BY license)

Setting aside the potential flaws of the axiomatic approach (Bayne, 2018), one of its virtues is that it emphasizes that subjectively conscious experience has dimensions that resist reduction to anything else. IIT thus acknowledges the sting of the hard problem of consciousness and sets out to tackle it head on, unlike competing approaches that, at points, might seem to dismiss the hard problem's real bite or focus on *function* rather than phenomenology. The Global Neuronal Workspace Theory (GNWT; Dehaene, 2014; Baars et al., 2021) might be an example.

However, compared to GNWT, IIT faces one fundamental limitation: It fails to acknowledge the crucial roles of attention in generating phenomenally conscious experience and shaping its contents. In cognitive science, attention is typically characterized as a process of selection and prioritization of a portion of all the information available to a cognitive system at a given time (Carrasco, 2018; Wu, 2024). Though the focus is typically on perceptual, top-down and voluntary selection, there is reason to believe that attention is a more general process of informational optimization, also encompassing cognitive, bottom-up and automatic varieties (Chun et al., 2011; Montemayor and Haladjian, 2015; Marchi, 2020; Lopez, 2022). Such pervasiveness of attentional processes increases the likelihood of them being implicated in different forms of consciousness (Marchetti, 2012; Pitts et al., 2018; Noah and Mangun, 2020).[1] A wealth of recent work in cognitive psychology, cognitive neuroscience, and evolutionary biology indeed indicates that, for animal consciousness (including of course humans), attention and consciousness cannot be fully dissociated (Haladjian and Montemayor, 2015; Montemayor and Haladjian, 2015). This evidence strongly suggests a partial dissociation, where though attention in its many forms is not sufficient for consciousness (hence the dissociation), some form of attention or other is necessary for any form of consciousness (Montemayor and Haladjian, 2015; Lopez, 2022).[2] This suggests that the project of giving a scientific account of phenomenal consciousness must involve an account of how phenomenal consciousness relates to attention.[3]

IIT's lack of an appropriate account of attention is not only a problem on the face of this evidence. It is also a problem by the theory's own lights. There is empirical and conceptual evidence that each of the countenanced essential properties of conscious experience involves attention in important ways (Kahneman, 1973; Merker, 2013; Watzl, 2014, 2017; Wiese, 2022; Marchetti, 2022). See Figure 2 for a schematic illustration.

---

[1] To be sure, the nature of attention is a matter of controversy in philosophy and cognitive science. It is precisely the variety of different mechanisms and processes associated with attention that has led some to be skeptical about there being a single thing deserving the name "attention", leading some to even ban the term from their labs (Rosenholtz, 2024). We grant that attention might be far from a monolithic thing, but we want to leave open the possibility that the diverse processes and mechanisms associated with attention share at least one interesting commonality, in virtue of which they can be considered "attentional" (for some examples see Fazekas and Nanay, 2021; Mole and Henry, 2023; Wu, 2024). The claims in this paper are compatible with the term "attention" ultimately referring to a family of different related processes.

[2] One immediate objection for this claim are dream experiences –we discuss them in section 3 below.

[3] We realize that we are taking a strong stance in the ongoing debate about the empirical dissociation between phenomenal and access consciousness (Block, 1995). As suggested by Sperling's (1960) well known study on iconic memory capacity and a wealth of subsequent research (a good recent example is Amir et al., 2023), it seems possible to have subjective experiences to which one lacks cognitive access. Since attention is connected to cognitive access (for instance, because it mediates the encoding of information into working memory), attention might seem to be necessary only for access but not for phenomenal consciousness. Our contention here is that some form of attention is also required for phenomenal consciousness, for instance along the lines suggested by Pitts et al. (2018). This view is further supported by recent studies showing that short-term memory capacities often thought to be pre-attentive (e.g. fragile visual short-term memory) are in fact also modulated by attention (see, e.g., Chiarella et al., 2023).

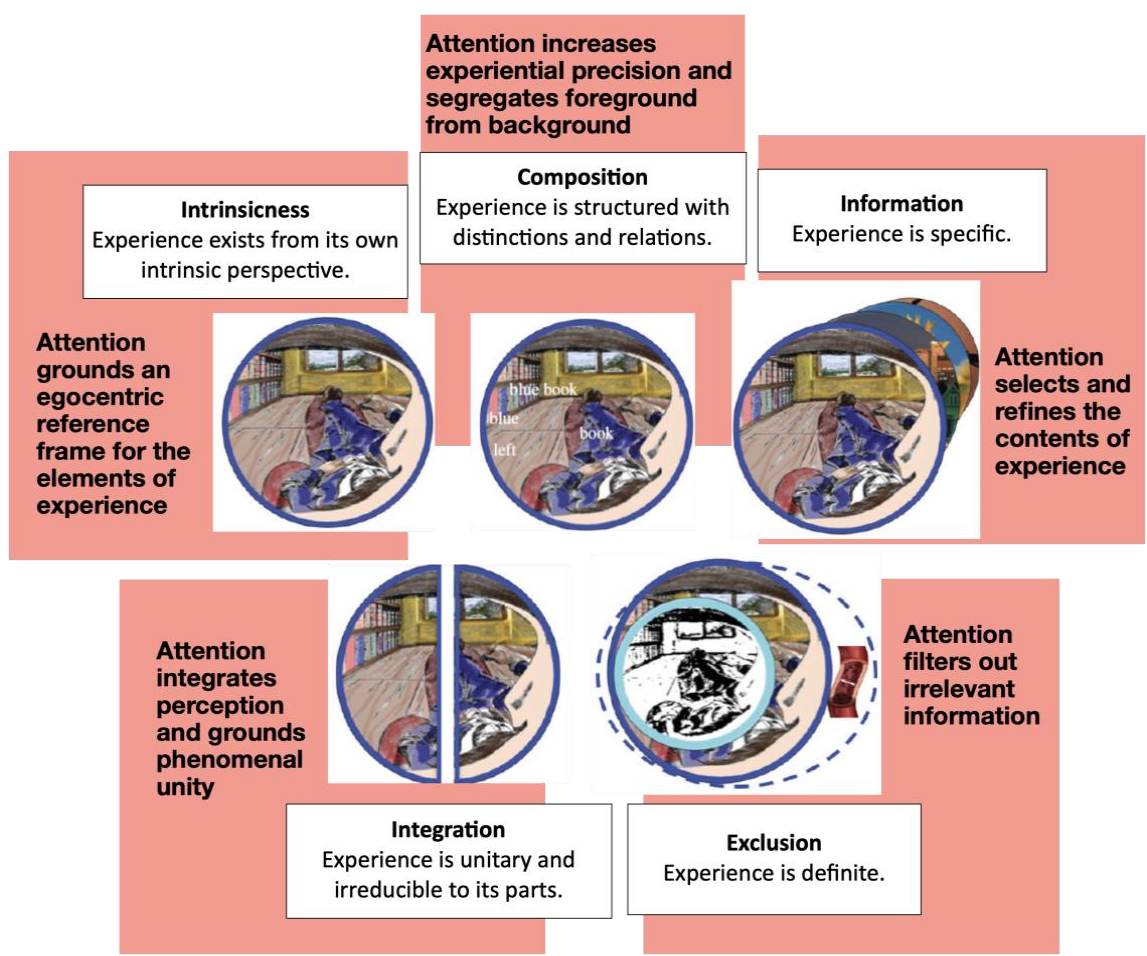

**Fig.2.** Attentional functions modulating properties of conscious experience.

Adapted from

https://en.m.wikipedia.org/wiki/File:Axioms_and_postulates_of_integrated_information_theory.jpg (CC BY license).

At present, IIT has no official claim about the relation between consciousness and attention (Koch, 2019: 204, n.16). However, some IIT proponents famously endorse a double dissociation (Koch, 2019; Koch and Tsuchiya, 2007). For instance, Koch (2019: 38) writes that *raw experience* can be dissociated from attention and other cognitive operations, and that selective attention is neither necessary nor sufficient for having conscious experiences. To be sure, however, Haun and Tononi (2019) recently offered what might be the first discussion of the role of attention within the IIT framework. Though this is a step in the right direction, we shall point at some important questions that this initial account must still address.

Our main goal in this paper is to show why IIT urgently needs an account of attention. The cited support for a partial dissociation and the pinpointed connections of the axiomatic phenomenal properties to attention provide good starting motivation for this idea but will not play the central role in our arguments. Instead, our arguments focus on the way IIT characterizes the physical substrate of consciousness, as a system that shall bear the essential properties of consciousness in virtue of its own intrinsic causal powers only. We thus start with a brief primer on four key claims of IIT (section 2). Next, we argue that these claims cannot explain important informational differences between different kinds of sensory experiences, and that this problem could be easily solved by invoking the roles of attention (section 3). We then argue that IIT is in fact incompatible with a double dissociation between consciousness and attention (section 4). We finish our discussion of IIT by discussing the seeds of an account of attention offered by Haun and Tononi (2019), while raising some initial questions (section 5).

Though our discussion is centered on IIT, the issues we raise potentially generalize to several other theories of consciousness. Specifically, IIT has much in common with primitivist (e.g. Pautz, 2009) and related internalist approaches to phenomenal content in philosophy (e.g., Horgan and Tienson, 2002; Mendelovici, 2018), as it emphasizes that the content of conscious experience is intrinsically determined with substantial independence from environmental conditions.[4] IIT could indeed be seen as a scientific formalization of such philosophical theories or at least some of their key claims, and thus it should be of interest to proponents of these views.[5] Relatedly, as we will see, IIT also has much in common with the structuralist approaches recently becoming popular in philosophy and cognitive science (Rosenthal, 2010; 2015; Northoff and Lamme, 2020; Kob, 2023; Fink and Kob, 2024, Kleiner, 2024). Thus, our arguments could potentially reveal the need for these other views to be explicit about the roles that attention plays in determining phenomenal (and, more generally, mental) content. We discuss our arguments' implications for these other theories in the final section of this paper (section 6). Overall, the most general take-home message of our discussion is that considerations about attention are indispensable for scientific as well as philosophical theorizing about conscious experience.

## 2      A quick IIT primer

We assume that IIT theorists, in line with the dominant trend in cognitive science, accept that the human brain is an information processing system.[6] According to IIT, the key processing difference between information processed unconsciously and information that becomes the content of a conscious experience is *integration*. Thus, the physical substrate of consciousness (PSC) in the human brain, that

---

[4] Internalism is the view that the contents of mental states in general are mostly and primarily determined by factors internal to the individual's mind and brain, rather than by the environment or other external factors. Internalist views of *phenomenal* content hold that these contents are primarily determined by their phenomenal qualities. These internalist views contrast with views for which environmental relations are crucial for determining phenomenal content; a prominent example is representationalism (Drestke, 1995). Primitivism (Pautz, 2009) and the phenomenal intentionality approach (Horgan and Tienson 2002; Mendelovici, 2018) are notable varieties of phenomenal internalism. See section 6 for more discussion of these views.
[5] An example is Pautz (2019).
[6] Though IIT does not in principle commit to any specific scale for the relevant units, for operational purposes proponents have recently focused on neurons or neuronal populations (Melloni et al., 2023). Here we follow this practice.

is, the neural mechanisms that give raise to conscious experience, should be those that maximize *informational integration*. Though this is the claim typically associated with IIT, IIT theorists have recently emphasized that the relevant information must be *intrinsic*. With this, they set their own sense of "information" apart from "extrinsic" kinds of information, such as Shannon information or information as the content of a message (Tononi, Boly et al., 2022; Mudrik et al., 2014). Thus, IIT's main claim is better captured thus:

**IIT1** The PSC in the human brain is the neural complex that maximizes intrinsic integrated information.[7]

To be sure, IIT has not been too clear about what "information" means. In fact, the term's meaning seems to have been changing since the first theory iteration. In the most recent version, IIT 4.0 (Albantakis et al., 2023), "information" is still not defined on its own but is instead implicitly characterized in connection with the notions of intrinsicality and integration. According to the postulate of information:

> *Intrinsic information* = "a measure of the *difference* a system takes and makes over itself for a given cause state and effect state" (Albantakis et al., 2023: 5).

In turn, according to the postulate of integration:

> *Integrated information* = a measure of the *irreducibility* of the cause-effect state of a whole set of units to separate subsets of units (Albantakis et al., 2023: 5).

Evidently, the sense of "information" at play is understood in causal terms. It remains to be seen to what extent does such a kind of information come apart, as intended, from extrinsic or message-like notions; however, this is not our current main concern. More important for our present purposes is that IIT endorses something resembling a phase transition between information in this widespread extrinsic sense and information in the countenanced intrinsic sense. We shall come back to this point in due course. Also important is that, due to this emphasis on the intrinsicality of information, IIT is committed to a strong kind of primitivism about conscious contents. At the very least, IIT is committed to a form of internalism. Such commitments will be important when drawing the implications of our arguments for theories of consciousness beyond IIT. But coming back for now to the relevant notion of information, here is the characterization afforded by the cited postulates of information and integration:

**IIT2** A neural complex N has intrinsic integrated information iff N has a *cause-effect structure* such that:
  (i) the *cause-effect state* of N is *irreducible* to states of subsets of N, and

---

[7] IIT theorists also emphasize that the five axioms should be taken together; thus, an exhaustive characterization of their main claim should also incorporate claims about composition and exclusion.

      (ii)      in this cause-effect state, N *takes and makes a difference over itself*.

Crucially, the key notions of *cause-effect structure*, *cause-effect state*, *irreducibility, and taking and making a difference over oneself* are all mathematically defined. The irreducibility of integrated information is a property measured by a difference between the causal interactions of the whole system and those of the minimal system subset.[8] This difference is indexed by IIT's well-known measure, $\Phi$: the greater the difference, the greater the $\Phi$ score. In addition, IIT 4.0 incorporates a new formal measure for intrinsic information, *ii*. This measure indexes the *causal impact* of the system on itself, resulting from two factors, *selectivity* and *informativeness*, which are also mathematically defined. They concern, respectively, the amount of uncertainty that the system is in one state rather than another, and the amount of deviation from chance in system state transitions (Albantakis et al., 2023: 15). Intrinsic information thus measures a property inversely correlated with both uncertainty and deviation from chance: the more uncertainty and deviation from chance, the less intrinsic information, and vice versa.

Intrinsic integrated information, as characterized in IIT2, is a necessary but not a sufficient condition for conscious contents. As IIT1 states, a neural complex is a PSC only if it *maximizes* intrinsic integrated information. This means that said complex must have *high* $\Phi$ and *ii* scores. *How* high is an interesting question. There is no absolute $\Phi$ or *ii* threshold; rather, whether a complex has relevantly high scores is determined relatively to the scores of the complex's components:

**IIT3**    A neural complex N maximizes intrinsic integrated information iff N's $\Phi$ and *ii* scores are *higher than* the $\Phi$ and *ii* scores of N's subsets.

To be sure, IIT1–3 are claims about the *presence* (or *degree*) of consciousness. Since IIT is offered as a general theory of consciousness, these claims shall apply to all varieties of conscious experience, including perception, illusion, hallucination, and dreams. However, IIT also makes an often-overlooked claim about the *quality* of conscious experience:

**IIT4**    A neural complex N supports conscious state C rather than C' iff the causal relations within N have structure S rather than S'.

IIT4 states *why* a neural state should give raise to a specific content conscious content with a specific qualitative feel. In this way, while IIT1–3 state what different experiences of different types have in common, IIT4 underpins the phenomenal differences between different types of experiences, e.g. perceptual experiences vs. dreams, as well as differences between specific experiential contents. Since such differences are explained in terms of differences in structures, IIT shares much with structuralist approaches to consciousness, both ontic (i.e., identifying phenomenally conscious experiences with structures; Rosenthal, 2010; 2015) and epistemic (i.e., using structure as a guide into the neural correlates of consciousness; Kob 2023, Fink and Kob 2024; see also Northoff and Lamme 2020). As with primitivism, this similarity will be important when drawing the overarching implications of our present criticisms to IIT.

---

[8] The lack of an independent definition of information gives this characterization a painful tinge of circularity, as the relevant difference is supposed to be an *informational* difference: between information in the whole system and information in the *less informative* system subset, that is, the system subset with the lesser amount of information.

## 3      The problem of informational differences

It is an axiom of IIT that conscious experience is maximally specific and detailed (axiom of information; see Figure 1). Conscious experience is "specific rather than generic" (Ellia et al., 2021: 5), it is "the way it is" and no other way (Albantakis et al., 2023: 5), and it is always "this one" (Tononi, Albantakis et al., 2022: 4). The new measure of intrinsic information, *ii*, shall be the most direct index of these properties of experience since it is introduced as a part of the postulate of information. Hence, the intrinsicality of information and the specificity of phenomenal content are practically equivalent in IIT: the specificity and detail of content of conscious experience is equivalent to the quantity of intrinsic information in the PSC, as indexed by the *ii* score. Since the PSC maximizes intrinsic information, phenomenal contents are the most specific and thus informative contents available in the brain, as opposed to the contents of unconscious representations.

The problem with this link between the specificity of phenomenal contents and intrinsicality of information is that, on the face of it, it cannot capture the fact that some types of experiences appear to be *more informative* than others. Take the paradigmatic case of conscious perception. Plausibly, conscious perception is maximally specific and informative because it provides maximally determinate contents, compared to, say, imaginative and dream experiences. Plausibly, the reason why conscious perception is maximally specific and informative is that it connects the experiencer with the environment, which supplies rich and abundant information about specific objects and properties. However, according to the high *ii* requirement, only information within the PSC is relevant for consciousness. Environmental information supplied by perception is not (or at least, not clearly) information *within the PSC*, that is, it does not (or not clearly) concern the causal impact of the PSC on itself. On this, Albantakis et al. (2023: 35) say that "the Φ-structure of a complex depends on the causal interactions between system subsets, not on the system interaction with its environment". To justify this, they emphasize their anti-functionalist stance, which is concerned with what consciousness *is*, rather than what consciousness *does*. Interactions with the environment are conceived as input-output functions, which can be the same even in systems with different internal causal structures. Only the latter shall matter for consciousness. Hence, IIT poses a discontinuity between the external channels of information and the intrinsic information that determines the maximally informative phenomenal content, because the only relevant structure of informational content that matters is intrinsic to the PSC.

This is already an odd view, because how could environmental information be irrelevant for perceptual experiences? Moreover, this issue is also of significance to a host of problems in philosophy of perception: the distinction between hallucinations and perception, the argument from hallucination, and the relation between narrow and wide content. For IIT, these are problems that must be solved in terms of the measures of information it postulates, the *ii* and $\Phi$ metrics. "Outside" information is not relevant, only *intrinsic* information counts. But from here, it is unclear how to distinguish perception from, say, dreams, in terms of informativeness. Perceptual consciousness is more informative than dream consciousness because the details of its contents concern aspects of the environment; in turn, false precision or acuity, as seen in dreams, is disinformation, and it should bring the *ii* measure considerably down. Put differently, the kind of informational mapping required for perceptual informativeness is much less trivial that the one required for dreams or hallucinations, which can occur in the absence of accuracy –the consequence of being wrong in a dream is to wake up from a nightmare, the consequences of being wrong about your surroundings can range from not ideal to truly catastrophic. But this should mean, contrary to the intrinsic sufficiency of the PSC, that external matching conditions are crucial to determine informativeness.[9]

Note that our concern with informativeness differences is not about whether the dreamer is aware that they are dreaming. Dream experiences are less informative than perceptual experiences not because the dreamer knows that what they are experiencing is a dream detached from their environment. Rather, dream experiences are less informative because the online experience that the dreamer is having need not contain the amount of detail that an analogous perceptual experience would. Consider standing in front of a menacing tiger in waking life versus in a dream. In waking life, you can see how the tiger's paws are positioned, the angle of its head, whether its coat is more or less reddish, etc. In a dream, these pieces of information can be more indeterminate. Also, in waking life, you can make some guesses about how the tiger got there, as well as predict likely outcomes. In dreams, these causal paths are also more indeterminate, and the possibility space is broader and less constrained. It is these differences that must be explained by IIT, and this in purely intrinsic terms, according to their own desiderata.[10]

As far as we can see, IIT could offer two responses. One response would be denying that conscious perception and dreams differ with respect to the relevant kind of information. Sure, my visual experience of my mother's face when she is sitting directly in front of me carries much more specificity and detail than her image in my dreams. However, the way how my visual percept and my dream image appear to me as "this one", "this very experience", need not differ. Alternatively, differences in structure (see IIT4) could be invoked to explain why perceptual experiences *feel more informative* than dream imagery. It could be that the PSC has a less complex causal structure during dreams than during wakeful conscious perception. This would mean that the latter involves many more distinctions and relations than the former (axiom of composition; Albantakis et al., 2023: 4).[11]

---

[9] That said, there is an interesting sense in which dreams may be informative with respect to the external environment. According to the Threat Simulation Theory (TST; Valli and Revonsuo, 2009), dreaming evolved as an offline simulation of the perceptual world, which could help with threat recognition and avoidance behaviors in waking life. This is an important sense in which dreams can be said to improve our information about the environment, but it is different from the sense we describe in the main text, which intends to stay close to IIT's notion of intrinsic information. We thank a reviewer from this journal for bringing this line of thought to our attention.

[10] Note, however, that if the dreamer became aware that they are dreaming, as it happens in lucid dreams, this should raise the $\Phi$-score of the dream experience, plausibly also raising the *ii*-score, due to the addition of the information that the experience is a dream.

The first response has some appeal, but it demands a significant shift in our understanding of what makes conscious experience specific and determinate. Differences in structure between different kinds of experiences seem to be the most promising way to go. However, it will still be very hard to explain the differences in the number of distinctions and relations without appealing to online environmental input.

Here is where attention comes in. Attention seems really crucial in providing the mapping that secures that perceptual informativeness correlates with behavioral success. As mentioned in the introduction, attention is linked to the selection and prioritization of a portion of the information available to a cognitive system at a given time (Carrasco, 2018; Wu, 2024). This includes, prominently, perceptual information. Thus, the proposed view is as follows. In typical perceptual experience, externally directed attention facilitates that environmental features and feature configurations are built into phenomenal distinctions and relations (as posited in the axioms of information and composition). The features in your focus of attention will appear in your experience with special specificity and detail (Nanay, 2010; Stazicker, 2011; Brogaard, 2015; though see Lopez and Simsova, 2023). Contrastingly, during dream experiences, attention is internally directed (Chun et al., 2011) and it is likely more diffuse, in the sense that it might not have such a well-defined focus.

Admittedly, here one might object that dream experiences in fact look like a very good example of conscious experience without attention. However, we can offer two reasons to reconsider this claim. First, while it is plausible that dreams do not involve attention if one thinks of attention as essentially linked to the performance of a task (for instance, we see such proposals in Rosenholtz, 2024 and Wu, 2024), there are other conceptualizations of attention that make it more plausible for it to be present in dreams; for instance, attention as a mechanism for availability of intermediate level representations (Prinz, 2012), or attention as a structure of the phenomenal field (Watzl, 2017; Jennings, 2020). Second, some attention-based models of dreams are already available. For instance, Conduit et al. (2000) argue that brain waves during REM sleep represent enhanced arousal of attention mechanisms. Relatedly, Solomonova and Carr (2022) argue that there is attention *in* dreams and that this attention mediates dream experience.

---

[11] We say more about these notions in section 5.

To summarize, our worry is that high amounts of intrinsic integrated information do not seem to correspond with the kind of informativeness we find in conscious experience, because the links with *external* information, deemed irrelevant for the Φ and *ii* measures, are in fact the best way of explaining why some kinds of experience (e.g., conscious perception) seem more informative and specific than others (e.g., dreams). This lack of correspondence is especially worrisome because IIT wants to explain the very specific way a phenomenal content feels in terms of the structure of causal relations yielding high Φ and *ii* scores, but the needed specificity seems to importantly depend on appropriate environmental connections. We propose that attention is crucial for establishing such connections. As we suggest in the next section, attention shapes the informational boundary of the PSC, operating like a valve that determines which and how much information makes it in. Indeed, as we also argue below, IIT seems to entail this. But then, attention to environmental features would be at least partly responsible for the PSC's Φ and *ii* scores, and hence these would not be entirely intrinsic to the PSC. Furthermore, these considerations also suggest that a double dissociation between attention and consciousness is untenable by IIT's own lights.

One final point on the relation between attention and intrinsic information. Since intrinsic information is defined as the product of *selectivity* and *informativeness*, a high *ii* score can be due to high selectivity, high informativeness, or high selectivity *and* informativeness.[12] Now, at least one of these factors, namely selectivity, is clearly related to attention. This is not only because of attention's widely acknowledged link with selection. In addition, a connection is suggested by the way selectivity is defined. As mentioned above, selectivity is a matter of the amount of uncertainty about the system's state: the lower the uncertainty, the higher the selectivity. Albantakis et al. (2023) say that uncertainty tends to increase with complex size, as larger complexes have more cause-effect states to "select from".[13] If, as we are about to argue, attention shapes the boundary of the PSC, then attention in fact contributes to increasing its *ii* score by decreasing uncertainty and thus increasing selectivity.

## 4    A double dissociation?

The core IIT literature does not make any reference to attention. Since IIT is fundamentally a mathematical theory, the need for it to invoke other cognitive capacities is not obvious. However, from this it does not follow that attention is irrelevant for the theory's goals, or that attention is (doubly!) dissociated from consciousness. We will now argue that, against these ideas, IIT is in fact incompatible with a double dissociation.

---

[12] Intriguingly, the two properties are in tension: the factors that tend to make a system more selective/less uncertain (e.g., system size) also make the system less informative/deviate more from chance (Albantakis et al., 2023: 15). See more discussion in section 5.
[13] See note 22 below.

IIT makes it explicit that there is a boundary, specified as the informational limit of the PSC with respect to the rest of the neurons in the brain, that constrains the amount of information that gets into the PSC (Tononi, Albantakis et al., 2022: 5). This boundary separates the neural complex that maximizes intrinsic integrated information from non-maximizing complexes. We have seen that the amount of intrinsic integrated information in a system is a matter of its causal powers over itself (*ii*) plus the irreducibility of these causal powers (Φ). Figure 3 illustrates these notions. Complexes 1 and 2 comprise the same number of units, but they have different amounts of intrinsic integrated information. One key difference is that the total causal state of Complex 1 at any given time is reducible to the causal states of sub-complexes AB and CD, but the total causal state of Complex 2 is not thus reducible. This gives Complex 2 a much higher Φ score.

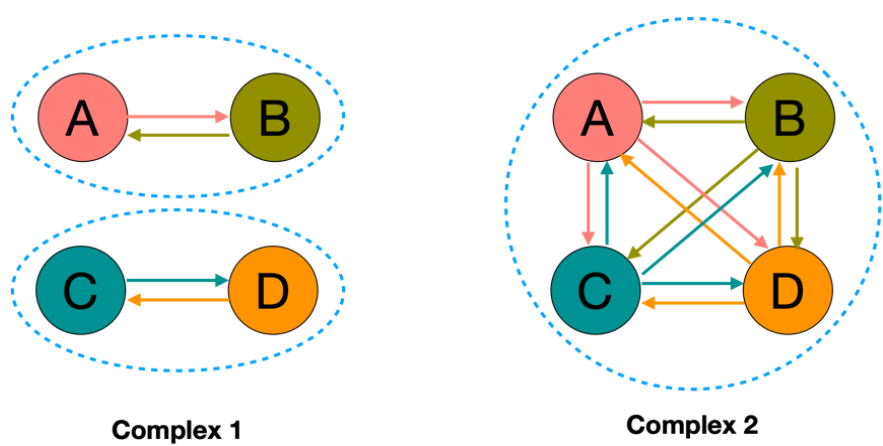

**Fig.3.** Two complexes with different amounts of informational integration
Source: Authors

Complexes 1 and 2 also have different informational boundaries (represented by the dotted lines). In Complex 1, each of AB and CD has higher Φ than ABCD. On the principle that the PSC must be the complex maximizing Φ, the informational boundaries of Complex 1 exclude ABCD as a candidate PSC. Contrastingly, the informational boundary of Complex 2 excludes AB and CD as candidate PSCs, as ABCD's Φ score is higher.

The notion of an informational boundary thus captures the key idea of IIT's postulate of exclusion, namely, that within a set of units, only the complex with the highest Φ bears consciousness. In the vivid language of Tononi, Albantakis et al. (2022), excluded complexes are mere informational "dust", as their Φ score is negligible, compared to the PSC's. Notably, this means that there is a big ontological jump between information within and outside the PSC. AB and CD, which on their own had an amount of integrated information that made them qualify as PSCs and bear a degree of consciousness (see IIT3), lose their claim as PSCs by becoming integrated within ABCD.

The extinction or exclusion of regions with non-maximal $\Phi$ as causally and meaningfully irrelevant even though they have a (potentially high) degree of $\Phi$ is indeed a feature of IIT that deserves closer examination. A striking aspect of this exclusion is that it introduces a binary division: either the information is causally and semantically relevant (it is within the PSC), or it is neither causally nor semantically relevant (it is "dust"). This is the sense in which IIT seems committed a phase transition from unconscious to conscious states. But on the other hand, both $\Phi$ and *ii* are magnitudes or analog measures, continuously indexing the degrees of integration and intrinsicality. Thus, information has more or less degrees of $\Phi$ and *ii*. This clearly stands in opposition to the binary exclusion principle, according to which information is maximally relevant or completely irrelevant. One way of reconciling this discrepancy is by appealing to varying thresholds for a sharp transition, as in "winner-takes-all" strategies. But this needs to be more explicitly addressed by IIT and clearly attention would play a key role in how this strategy is implemented.

As it stands, then, IIT faces a tension within its core tenets. Endorsing a double dissociation between consciousness and attention would make matters worse, as attention provides a promising way to explain away this tension. IIT could emphasize the selective function of attention and conceptualize it as an informational valve that determines the size of the PSC by regulating and organizing the flow of information. Attention would then be responsible for "eliminating" complexes with less than maximal $\Phi$ and *ii* and would thus play an essential role in explaining the binary phase transition between unconscious information processing outside the PSC and intrinsic conscious information within the PSC. This might be a problem for theories of consciousness in general, but because of its commitments to information flow and the sharp boundary of the conscious substrate, it is particularly evident for IIT.

While this is already motivation for IIT *not* to endorse a double dissociation, there is also reason to think that such dissociation is untenable for IIT. In the brain, the mappings of informational interaction with neural activity that gets excluded or turned into "dust" contain the information that is either modularly processed or processed by neural complexes with low $\Phi$ score. This means that high levels of integration and causal structure depend on the initial information of the relevant parts of the brain that shape the border of the PSC. Now, according to some IIT proponents, attention is responsible for early processing, and attention can process information unconsciously because it is doubly dissociated from phenomenal consciousness (Koch, 2019; Koch and Tsuchiya, 2007). This means that attentional processing is dissociated from information within the PSC, which has the maximum amount of independence in the brain, and also that consciousness can operate independently from attention. But this clearly cannot mean that the PSC can simply free float within a sea of information in the brain. Since attention processes the first stages of information, it is crucial to determine all the information that constitutes the boundary of the PSC. Consequently, the information coming from attention shapes the PSC, both in content and size.

So we get the following picture. Attention processes information in a modular-like fashion. Most of these processes occur unconsciously. However, once there is a substrate with maximal $\Phi$, all the information that was processed unconsciously by attention achieves a new status because the information is now integrated into a maximal cohesive unit with causal powers, rather than mere pieces of information processed for specific goals. The PSC has full informational and causal independence, which is irreducible to the subcomponents that processed information at early stages. But the information at the boundary of the PSC is all provided by areas of the brain with lower $\Phi$. Therefore, attention is necessary for the boundary of the PSC to have the information it does. Thus, IIT must acknowledge attention as necessary (even if not sufficient) for consciousness.[14]

This consequence clearly is incompatible with the claim that attention is doubly dissociated from consciousness, even though it is still compatible with a single dissociation –as attention can operate without consciousness. More precisely, while attention can operate with full independence in "algorithmically" processing information for specific goals (Haladjian and Montemayor, 2023), consciousness depends on attention to specify a boundary of information that shapes the contents and causal powers of the region with maximal $\Phi$. Therefore, the dissociation is asymmetric and not double: consciousness depends on attention, but attention does not depend on consciousness.

This makes a lot of sense from an evolutionary and theoretical point of view. But our key point here is that IIT creates a tension within its own conceptualization of the PSC. First, without attention the PSC could be conceived as a deeply solitary or solipsistic entity, with a boundary shaped by attention, but somehow purely independent of the "informational dust" surrounding it. Second, how the boundary is shaped presumably determines the size of the PSC every time a maximum of $\Phi$ is reached, including during dreams. But, as we saw, IIT needs to make sense of the distinction between perceptual and dream experiences. Proponents of IIT could say that they are not concerned with distinguishing perception from dreams, but we believe that explaining this distinction is a minimum constraint on any view of consciousness because we want to preserve the difference between consciousness with and without interaction with the world. Moreover, as discussed above, providing an experience-environment mapping that secures a correlation between perceptual informativeness and behavioral success is a more pressing issue for IIT than for many other theories of consciousness, because of its emphasis on the intrinsicality of information.

---

[14] Pitts et al. (2018: 6) make a similar point: "It may be the case that attention plays a crucial role in determining the shape of the structure of integrated information (what is in versus out of the major complex), and therefore the content that we consciously experience, while the more basic distinction between experience and no-experience may not depend on attention. Alternatively, these two aspects of consciousness [i.e., presence vs absence and contents/quality] may be intimately linked, because a common way to distinguish conscious from unconscious states is to assess whether any contents can be consciously experienced."

Finally, if attention is an informational valve that determines the size of the flow of information into the PSC and how the contents of the PSC are shaped, then it needs to be clearer how the mappings at the boundary differ from mappings inside the PSC, given that, presumably, it should be possible to obtain all the information that gets into the PSC from the information at the boundary. This calls into question the full informational independence of phenomenal contents from attention.

## 5 IIT on spatial attention

Though explicit discussion of attention is scarce in the IIT literature, in a recent article Haun and Tononi (2019) acknowledge the effects of attention on spatial experiences and propose an account of these effects within IIT's framework, with the aid of a response gain model of spatial attention.[15] For the reasons discussed above, we think that this is an important addition to the theory that deserves further elaboration.

Haun and Tononi focus on *spatial* attention and how it modulates the visual experience of looking at an empty, boundless canvas. This choice of example shall illustrate the experience of visual space, arguably a pervasive and easily introspectable component of everyday experience. As usual in IIT, they start from considering the phenomenological effects of spatial attention, and from there they move to consider which physical mechanisms could bring about and sustain such phenomenological effects. In the light of William James' (1890) celebrated characterization of attention as a "a concentration and focalization of consciousness" that "implies withdrawal from some things, in order to deal effectively with others", along with Intriligator and Cavanagh's (2001) research on the windows of spatial attention, they pinpoint two phenomenological marks of spatial attention (Haun and Tononi, 2019: 23):

- *Flexible Spotlight*: The "spotlight" of attention can shift from one location to another, as well as expand or contract.
- *Highlighting:* The attended region of space appears to be "highlighted", as if more of the experience were concentrated there, *at the expense of the rest of the canvas*.

To illustrate, suppose you are looking at the blue sky in a clear spring afternoon, and suppose you can identify eight portions in this visual experience (see Figure 4; see also figure 1 in Haun & Tononi, 2019). Without moving your head or eyes, that is, without changing your sensory input, you could switch your attention around from A to B, B to C, etc. In doing this, the quality of your experience changes: first A is

---

[15] The spotlight view can still work for spatial attention; however, in the light of non-spatial forms of attention, (e.g., object or feature attention), it is no longer accepted as a general account of attention. Haun and Tononi's account then requires further elaboration also in this respect.

"highlighted", then B, etc.[16] Similarly, you could focus attention on A alone, on left-side half ABCD, on upper left quadrant AC, etc., so that your attentional "spotlight" expands or contracts.

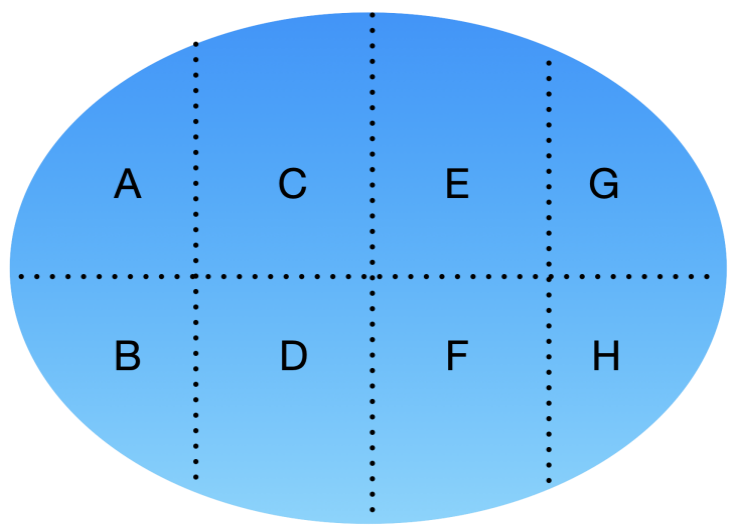

**Fig.4**. Experiencing empty space.
Source: Authors

*Flexible Spotlight* and *Highlighting* describe innocuous phenomenological data, but they reveal one potential conceptual tension for IIT. Suppose that you focus your attention on ABCD. ABCD would then be "highlighted", as if more of the experience was concentrated in ABCD *at the expense of* EFGH. This could mean two things for EFGH: either your experience of EFGH is weakened or de-emphasized (the opposite of highlighting), or it fades out altogether. Given IIT's assumption that conscious experience does not depend on attention, what is meant in *Highlighting* is probably the first option, so that EFGH is still experienced in some way –so "less of the experience" is concentrated in EFGH. It might be helpful to list these points schematically:

- **(A1)** A non-highlighted portion $E_1$ of experience E is still a portion of E (it does not fade out from E).
- **(A2)** A non-highlighted portion $E_1$ of experience E "concentrates less of E" than a highlighted portion $E_2$.

Now, say that "concentrating experience" involves having more of what makes up experience, just like juice concentrate has more of what makes up juice –arguably.[17] For IIT, what makes up experience are the causal properties indexed by $\Phi$ and/or *ii*. Hence, attention shall affect these causal properties, so that the Φ and/or *ii* scores of ABCD's substrate increase. For this to occur *at the expense* of other experiential contents, like EFGH, then the Φ and/or *ii* scores of EFGH's substrate should decrease.

---

[16] Carrasco et al. (2004) offer a seminal experimental demonstration of this phenomenal effect.
[17] We shall not make any claims about the metaphysics of juice!

**(A3)** If $E_2$ is highlighted and $E_1$ is not, the Φ/*ii* scores of $E_1$'s substrate decrease with respect to the Φ/*ii* scores of $E_2$'s substrate.

However, as we have seen, IIT also claims that at any given time, only one complex within a single set of units can support consciousness, namely, the complex with the higher Φ and *ii* scores (recall how complex ABCD in Figure 3 "excludes" complexes AB and CD). All other complexes not only do not support consciousness, but also are "dust" and "excluded from existence" (Tononi, Albantakis et al., 2022: 7-8). Thus, there is a strong implication that attending to ABCD extinguishes EFGH.

**(A4)** If the substrates of $E_1$ and $E_2$ are part of the same complex (the substrate of E), and the substrate of $E_1$ has higher Φ/*ii* scores than the substrate of $E_2$, then only the substrate of $E_1$ bears consciousness.

Above, we suggested that it would be useful for IIT to incorporate attention as an informational valve that determines the boundary of the PSC, that is, its size and external shape. Our suggestion is now that the roles of attention in determining the *internal* structure of the PSC for particular experiences should be emphasized and further clarified. IIT must account for attentional highlighting without saying that when you highlight ABCD, EFGH disappears from your experience. However, to be consistent with IIT's core tenets, this account should still allow that highlighting ABCD boosts the Φ and *ii* scores of ABCD's substrate.

Consider now what the response gain model of spatial attention brings to IIT's table. In humans and other mammals, spatial attention increases the spiking rate per time unit of neurons tuned for the attended location (Reynolds and Heeger, 2009). Haun and Tononi simulate this effect by lowering the activation threshold of a subset of units within a grid. This is illustrated in Figure 5.[18] There, we see an 8-unit linear grid ABCDEFGH in an "all off" state under two conditions: with and without an "attentional spotlight" in units CDEF.[19] Without attention, all units have the same activation threshold. With an attentional spotlight in CDEF, CDEF have lower activation thresholds.

---

[18] Figures 5 and 6 are directly taken from Haun and Tononi's (2019) Figure 11. We present these as separate figures to facilitate exposition of the central ideas involved.

[19] As usual within IIT's framework, units have two possible states: On or Off. "All off" is thus one of 256 possible states of an 8-unit grid. These possible grid states are represented by the arrays of red dots in Figure 5; they are the same for the "unattended" and "attended" grids since they are both "All off".

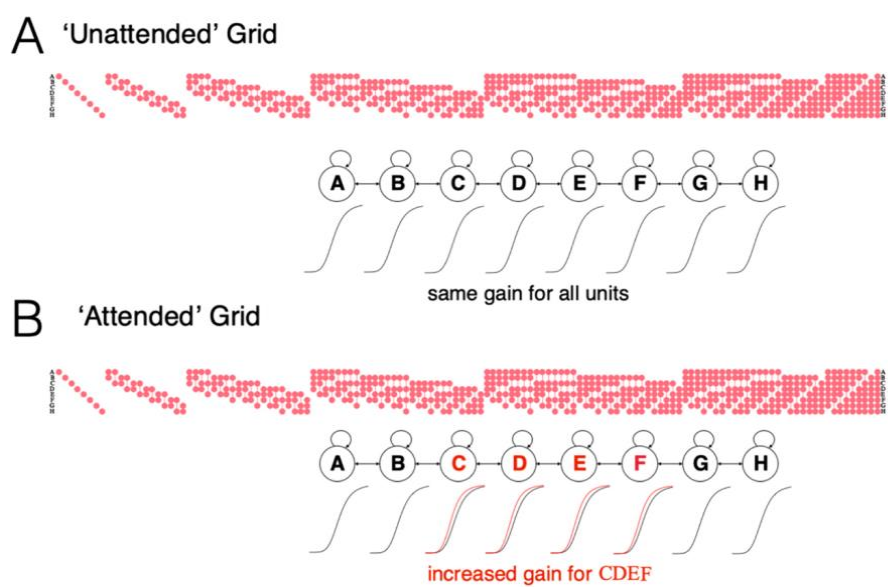

**Fig.5.** Spatial attention simulated as response gain.
Adapted from https://www.mdpi.com/1099-4300/21/12/1160 (Figure 11; CC BY license)

Notably, this response gain simulation leaves out another important aspect of spatial attention, namely, its suppression effects (Gouws et al., 2014). Attending to a spatial location involves actively suppressing information from other locations, so that cell responses to these other locations are inhibited. This aspect of the mechanism of spatial attention is of great significance for an account of what it means to experience attended portions of space as highlighted, *at the expense of others*. However, acknowledging such suppression within IIT's framework might put further pressure on the idea that one can experience attended portions of the visual field (as highlighted) together with unattended portions. If the substrates of the latter are actively suppressed, then its Φ/*ii* scores should go down, potentially extinguishing these experiential contents.

On the other hand, though, the links between response gain (and suppression) and Φ scores (and, eventually, also *ii* scores) are not clear. Specifically, it is not clear how attentional response gain affects informational integration, as measured by Φ, given that Φ underscores quantity of consciousness. If attention brings about something like a higher concentration of consciousness, then attention should increase Φ. But it is not clear how we get that from response gain. Compare once again complexes 1 and 2 in Figure 3. The reason Complex 2 has higher Φ than Complex 1 is that the number of recurrent connections amongst its elements is greater. But grids A and B in Figure 5 do not differ in number of recurrent connections.

Haun and Tononi note that the key differences between the grids with and without an attentional spotlight, which reflect the experience of attentional highlighting, are differences in what they call the *context* of a *compound distinction*. These are technical terms that (roughly) refer to the set of the unit subsets comprised by a complex of units (e.g., complex CDEF comprises C, D, E, F, CD, DE, CDE, etc.) and the causal relations binding each of these compounds with others (e.g., C as a cause and as an effect of CD, of CDE, etc.). They illustrate these differences as depicted in Figure 6.

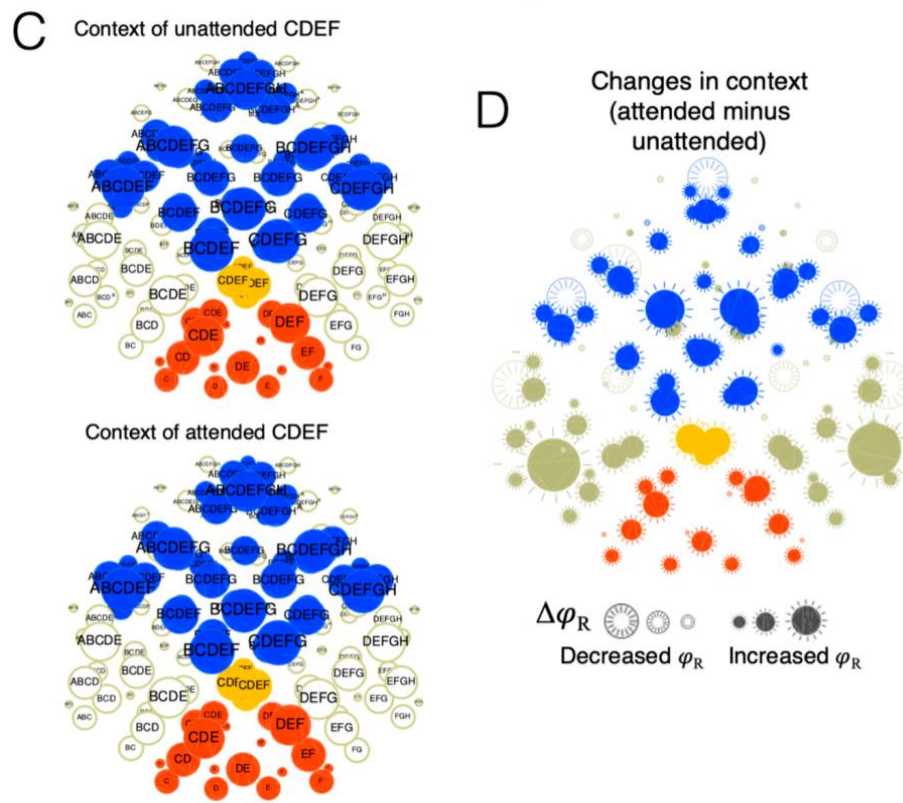

**Fig.6.** Attentional highlighting, according to IIT.
Source: https://www.mdpi.com/1099-4300/21/12/1160 (Figure 11; CC BY license)

Roughly, the plots on the left show the causal relations of CDEF with the rest of the ABCDEFGH structure with and without an attentional spotlight/response gain in CDEF. The plot on the right shows the difference between the two conditions in terms of differences in Φ-score for CDEF's different causal relations. Quite notably, Haun and Tononi (2019: 23) say that the relations between CDEF and the rest of the structure are *qualitatively similar* for both attended and unattended grids (Figure 6, left); however, they note that increasing the gain of CDEF involves "an enhancement of many relations", especially those involving CDEF, CDE, and DEF (Haun and Tononi, 2019: 21). By "enhancement", they seemingly mean an increase in Φ-scores. They say: "The context of CDEF is generally more irreducible when the gain of CDEF is increased by the 'spotlight' of attention" (Haun and Tononi 2019: 23).

We see the countenanced boost in irreducibility/Φ-score in the attended grid in that the scatterplot in the right side of Figure 6 has more solid circles than hollow ones.[20] Also, the preferential enhancement of CDEF, CDE and DEF relations is presumably evinced by the virtual lack of hollow circles for these compounds. Still, more clarity is needed on the countenanced transition between response gain as depicted in Figure 5, and attentional highlighting as depicted in Figure 6, given that attentional highlighting does not alter the number of recurrent connections. This point is also key for understanding why unattended portions of experience are not extinguished.

Haun and Tononi could also emphasize that, as we mentioned at the outset, Φ more readily indexes the presence or quantity of consciousness rather than its quality. Thus, in principle, two experiences with the same Φ score could still be qualitatively different due to differences in internal structure.[21] Then, if attentional highlighting is a matter of the quality, not quantity of consciousness, and as such, it is not a matter of increasing or decreasing Φ, our worry about extinguished portions of experience is misplaced. The point is well taken, but then we would invite Haun and Tononi to go beyond the somewhat metaphorical Jamesian characterization of the effects of attention as a concentration of consciousness. We think that this project should involve two things: looking at a more general conceptualization of attention as a process of informational enhancement or optimization (Marchi, 2020; Lopez, 2022), and elaborating on the connections between attention and intrinsic information, *ii*.

Regarding the connection between attention and *ii*, the analysis of this measure in terms of selectiveness and informativeness echoes our proposed conceptualization of attention as an *informational valve*. Selectivity and informativeness pull a complex in different directions, bringing about a tension between "expansion" and "dilution" (Albantakis et al., 2023: 9). Selectivity is supposed to be greater for smaller complexes, as these have fewer cause-effect states to "select from".[22] In turn,

---

[20] Roughly, each circle represents a compound within the ABCDEFGH grid. Colors represent different kinds of mereo-topological relations: open brown circles for *connection*, blue for *inclusion up*, orange for *inclusion down*, and yellow for *self-inclusion*. We take this key directly from Haun and Tononi (2019: 21).

[21] This move could address a keen objection raised by a reviewer from this journal, namely, that some changes in attention do not seem to come with changes in consciousness or conscious contents. A case in point are changes in feature-based attention. Consider an experiment by Andersen et al. (2008), where subjects see a cloud of swirling red and blue dots, and centrally there is a fixation cross which changes color between red and blue. Subjects are instructed to shift attention between the red and blue following color shifts in the fixation cross. One could argue that throughout these shifts of attention one retains a relatively stable experience of swirling dots and an alternating fixation color. However, there is reason to believe that feature-based attention does in fact alter the quality of experience. Recent studies show that feature-based attention modulates the contents of fragile visual short-term memory, a capacity that is associated with conscious experience by opponents of the view that consciousness requires attention (Chiarella et al., 2023). Within IIT's framework, one could accept that the internal structure of the experience and the involved causal relations changes, while the overall Φ-score remains unchanged..

[22] So far as we understand, the relevant states are the rows and columns in a complex's *transition probability matrix* (TPM; see Albantakis et al., 2023: 9). A TPM captures the complex's causal profile, showing how any given complex state increases or decreases the probability of the system's being in any other system state. Each row and column corresponds to one system state; row-column intersections show the probability that the system transitions from the one state to the other. A small system, comprised of few units, will in principle

informativeness shall be greater in larger complexes, as these "deviate more from chance" (Albantakis et al., 2023: 15).[23] Attention plausibly affects these properties when it modulates PSC size. And to the extent that these properties underpin "the difference that the system takes and makes over itself", it is plausible that the shifting balance of selectivity and informativeness underscore specific experienced contents moment to moment, including what is highlighted at the expense of what. This could be a promising way of specifying the effects of attention on experience at a much finer grain than just delimiting its external boundary from units that are not part of the PSC.

## 6      Implications for structuralist and primitivist theories of consciousness

We have raised some criticisms that make it evident that IIT needs an account of attention and of how it contributes to generating conscious experience and shaping its contents. We regard these criticisms as constructive, aiming at strengthening the theory's credentials as a scientific theory of consciousness and at making more palpable its applicability to the paradigmatic case of consciousness in human beings and other animals. We have also suggested possible paths for IIT to develop this account in consonance with the spirit of the theory.

We take the concerns raised here to be a manifestation of a general constraint on consciousness theorizing, namely: Consciousness theories must acknowledge and integrate the roles attention plays in determining many aspects of consciousness. Though it is not conclusively established that consciousness requires attention, evidence keeps accumulating in several fronts (Marchetti, 2012, 2022; Haladjian and Montemayor, 2015; Montemayor and Haladjian, 2015; Pitts et al., 2018; Noah and Mangun, 2020; Watzl, 2014, 2017). Consciousness theories cannot ignore attention. Though this is primarily a point about the science of consciousness, it shall plausibly apply to philosophical theorizing as well. To conclude this paper, we will briefly discuss the implications of our arguments for primitivist and structuralist approaches to consciousness and conscious contents.

We noted at the outset that, due to its emphasis on the intrinsicality of information, IIT seems to endorse an internalist position about conscious contents. Moreover, since phenomenal contents are structures that can only be triggered intrinsically, IIT also seems committed to a strong kind of primitivism, according to which the content and phenomenal character of experience depends exclusively on the structure of the PSC. Primitivism (e.g., Pautz, 2009) holds that phenomenal content cannot be explained in terms of something else, such as environmental relations. Primitivism is thus in overt opposition to "tracking" intentionalism (e.g., Dretske 1995), which is externalist through and through. In this respect,

---

have less rows and columns in this matrix. This gives the complex less states to 'select from', making it more selective by IIT's definition.

[23] Higher informativeness as increase deviation from chance amounts to greater values in the relevant TPM cells (see note 22).

primitivism could be susceptible to a related version of the problem of informational differences between perceptual experiences and imagistic or dream experiences.

A key difference between IIT and primitivist theories of consciousness is that the latter gain much of their appeal from an epistemic thesis, namely, that introspective access yields certainty with respect to consciously experienced contents, and that this is essential for the fulfillment of rational and doxastic roles. Though here we are not concerned with these epistemic roles, we concede that intrinsically determined phenomenal content may be rationally important, and that this remains a central appeal of primitivist views. However, primitivism must still tell a story on why perceptual attention seems to enrich experienced contents. Evidently, this story should not resort to how perceptual attention facilitates extracting environmental information.

Here we think that primitivism could take inspiration from IIT and sketch a solution in terms of how attention modifies the internal structure of experience. This takes us to the commonalities between IIT and structuralist views, as well as the consequences of our arguments for the latter (Kob 2023, Fink & Kob 2024; see also Northoff & Lamme 2020). Notably, on the face of it, structuralist views need not suffer from a severe disconnect from external information, as primitivism and IIT do. However, arguments recent offered by Lee (2021) suggest that attention shall also play a key role for structuralist approaches. Lee argues that structuralist theories should not model mental qualities as points in a quality space, but rather as *regions* in a quality space. According to Lee, this region-based conceptualization better captures the datum that conscious experiences (e.g., perceptual) admit of variations in precision. At the same time, the region-based conceptualization highlights the disconnect between the space of mental qualities and the space of perceptible qualities of external objects, as the latter does not admit of variations in precision. If this is the correct conceptualization of structuralism, then the need to acknowledge attention is clear, as attention is known for its role in modulating the precision of phenomenal contents (Nanay, 2010; Stazicker, 2011; Brogaard, 2015; though see Lopez & Simsova, 2023).[24]

Coming back to the applicability of our arguments to internalist theories, more significant implications concern what Horgan and Tienson (2002) call "separatist" views of intentionality or mental content. According to separatism, for mental states in general, phenomenal aspects are independent from intentional aspects. This entails that the phenomenal qualities of conscious experiences in principle come apart from their intentional contents, that is, the objects and properties represented by the experience. Our suggestion is that separatist views comport well with a double dissociation between attention and phenomenal consciousness, where attention is conceptualized as key for determining intentional contents, but these intentional contents are detachable from what the experience feels like.

---

[24] For dissent, see Block (2015). For a response, see Author Reference (forthcoming 2024).

This version of the article has been accepted for publication, after peer review but is not the Version of Record and does not reflect post-acceptance improvements, or any corrections. The Version of Record is available online at: https://doi.org/10.1007/s10670-025-00949-1

This reinforces our point that IIT is incompatible with a double dissociation. At least for conscious experiences, IIT seems to *oppose* separatism, since it conceptualizes the contents of conscious experience so that these are not independent and not separable from its phenomenal aspects; hence, IIT shall reject the double dissociation. But there is also a more general implication for internalist views. If, as we argued, attention is needed not only for fixing intentional content, but also for shaping the relations between internally determined phenomenal contents, then separatist views seem mistaken –if intentionality and phenomenality both depend on attention, then the two very likely hang together. This is very relevant for Horgan and Tienson's (2002) arguments against separatism and in favor of phenomenal intentionality, a kind of intentionality constitutively determined by phenomenology alone. Our arguments suggest that there is no kind of intentionality that is constitutively determined by phenomenology alone, with independence of environmental tracking.